\documentclass[11pt]{article}
\usepackage{times}
\usepackage[centertags]{amsmath}
\usepackage{mathrsfs,eucal, amssymb, amsthm}
\usepackage{natbib}

\usepackage[plain,noend]{algorithm2e}
\usepackage{comment, url, graphicx, graphics, bm}
\usepackage[final]{changes} %[final]
\def\o*{o_{{\!}_{P^*}}}
\def\O*{\cal O_{{\!}_{P^*}}}

\def\E{\mathrm{E}}
\def\Var{\mathrm{var}}

\def\ML{\mathrm{ML}}
\def\MHD{\mathrm{MHD}}
\def\AMHD{\mathrm{AMHD}}

\def\.{\mbox{.}}

\def\le{\leqslant}
\def\ge{\geqslant}
\def\sfrac(#1,#2){\mbox{$\frac{#1}{#2}$}}

\def\tr{{\mbox{\tiny{$\mathrm{T}$}}}}

\def\bernstein{Bernstein }

\def\ceiling#1{\left\lceil #1 \right\rceil}

\newtheorem{theorem}{Theorem}% [section]

\newtheorem{lemma}[theorem]{Lemma}

\newtheorem{remark}{Remark}[section]

\makeatletter
\renewcommand{\algocf@captiontext}[2]{#1\algocf@typo. \AlCapFnt{}#2} % text of caption
\def\@algocf@capt@plain{top}
\renewcommand{\algocf@makecaption}[2]{%
  \addtolength{\hsize}{\algomargin}%
  \sbox\@tempboxa{\algocf@captiontext{#1}{#2}}%
  \ifdim\wd\@tempboxa >\hsize%     % if caption is longer than a line
    \hskip .5\algomargin%
    \parbox[t]{\hsize}{\algocf@captiontext{#1}{#2}}% then caption is not centered
  \else%
    \global\@minipagefalse%
    \hbox to\hsize{\box\@tempboxa}% else caption is centered
  \fi%
  \addtolength{\hsize}{-\algomargin}%
}
\makeatother

\begin{document}
%\jname{Biometrika}
%%% The year, volume, and number are determined on publication
%\jyear{2012}
%\jvol{99}
%\jnum{1}
%%% The \doi{...} and \accessdate commands are used by the production team
%%\doi{10.1093/biomet/asm023}
%\accessdate{Advance Access publication on 31 July 2012}
%\copyrightinfo{\Copyright\ 2012 Biometrika Trust\goodbreak {\em Printed in Great Britain}}

%% These dates are usually set by the production team
%\received{April 2012}
%\revised{September 2012}

%% The left and right page headers are defined here:

\title%[Parametrization and Smoothing using \bernstein polynomials]
{\bernstein Polynomial Model for Grouped Continuous Data}
%\maketitle
%\begin{comment}
\author{Zhong Guan\\
Department of Mathematical Sciences\\ Indiana University South Bend\\ South Bend, IN 46634, USA\\
email: \texttt{zguan@iusb.edu}\\
}
\date{}
\maketitle
\begin{center}
\textbf{Abstract}
\end{center}
Grouped data are commonly encountered in applications. All data from  a continuous population are
grouped due to rounding of the individual observations.  The Bernstein polynomial  model  is proposed as an approximate model in this paper for estimating a univariate density function based on grouped data. The coefficients of the Bernstein polynomial, as the mixture proportions of beta distributions,  can be estimated using an EM algorithm. The optimal degree of the Bernstein polynomial can be determined using a change-point estimation method. The rate of convergence of the proposed density estimate to the true
 density is proved to be {\em almost parametric} by an acceptance-rejection arguments used in Monte Carlo method. The proposed method is compared with some existing methods in a simulation study  and is applied to the Chicken Embryo Data.

%\begin{keyword}[class=MSC]
%\kwd[Primary ]{62G07}
%\kwd{62G20}
%\kwd[; secondary ]{62E17}
%\end{keyword}


\section{Introduction}
In real world applications of statistics, many data are provided in the form of frequencies of observations in some fixed mutually exclusive intervals, which are called grouped data.
Strictly speaking, all the data from  a population  with a continuous distribution are
%truncated
grouped due to rounding of the individual observations \citep{Hall-1982-siam-j-app-math}.
The EM algorithm has been used to deal with grouped data \citep{Dempster1977}. \cite{McLachlan-Jones-1988-biom}
introduced the EM algorithm for fitting mixture model to grouped data \citep[see][also]{McLachlan-Jones-JAS-1990}.
Under a parametric model, let $f(x;\theta)$ be the probability density function (PDF) of the underlying distribution with an unknown parameter $\theta$.
The maximum likelihood estimate (MLE) of the parameter $\theta$ can be obtained from grouped data and is shown to be consistent and asymptotically normal \citep[see, for example, ][]{Lindley-1950-mpcps, Tallis-Technometrics-1967}. Parametric MLE is sensitive to model misspecification and outliers.  The minimum Hellinger distance  estimate (MHDE) %$\hat\theta_{\MHD}$
of the parameter using grouped continuous data
is both robust for contaminated data  and asymptotically efficient \citep{Beran-1977a-aos, Beran-1977b-aos}. Parametric methods for grouped data requires evaluating integrals which makes the computation expensive. To lower the computation cost \cite{Lin-and-He-2006-bka} proposed the approximate minimum Hellinger distance estimate (AMHDE) for grouped data by the data truncation and replacing the probabilities of class intervals with the first order Taylor expansion. Clearly their idea works for MLE based on grouped data.

Under nonparametric setting, the underlying PDF $f$ is unspecified.
Based on grouped data  $f$ can be estimated by the empirical density, the relative frequency distribution, which is actually a discrete probability mass function.  The kernel density estimation \citep{Rosenblatt-1956-ann-math-stat, Rosenblatt-1971}
can be applied to grouped data \citep[see][for example]{Linton-and-Whang-2002, Jang-and-Loh-2010-ann-appl-stat, Minoiu-and-Reddy-2014-J-Econ-Ineq}.
The effects of rounding, truncating, and grouping of the data on the kernel density estimate have been studied, maybe among others, by \cite{Hall-1982-siam-j-app-math}, \cite{scott-sheather-1985-comm-stat-theo-meth}, and \cite{Titterington-1983-comm-stat-theo-meth}. However, the expectation of kernel density estimate is the convolution
of $f$ and the kernel scaled by the bandwidth. It is crucial and difficult to select an appropriate bandwidth to balance between the bias and variance. Many authors have proposed different methods for data-based bandwidth selection
over the years. The readers are referred to a survey by  \cite{Jones-etal-1996} for details and more references therein. Another drawback of the kernel density is its boundary effect. Methods
of boundary-effect correction have been studied, among others,  by \cite{Rice-1984} and \cite{Jones-1993}.

``All models are wrong''\citep{Box-1976-JASA-Sci-and-Stat}. So all parametric models are subject to model misspecification.
The normal model is approximate because of the central limit theorem.
%However in most cases it is hard to access the closeness of the normal approximation to the real underlying distribution.
The goodness-of-fit tests and other methods for selecting a parametric model introduce additional errors to the statistical inference.

Any continuous function can be approximated by polynomials. \cite{Vitale1975} proposed to estimate the PDF $f$ by estimating the coefficients $f(i/m)$ of the \bernstein polynomial \citep{Bernstein} $\mathbb{B}f(t)=\sum_{i=0}^m f(i/m){m\choose i}t^i(1-t)^{m-i}$
by  $\hat f(i/m)=(m+1)\{F_n[(i+1)/(m+1)]-F_n[i/(m+1)]\}$, $i=0,1,\ldots,m$, where $F_n$ is the empirical distribution function of $x_1,\ldots,x_n$. Since then, many authors have applied the \bernstein polynomial in statistics in similar ways \citep[see][for more references]{Guan-arXiv-2014}.
These and the kernel methods are {\em not
model-based} and not maximum likelihood method. Thus they are not efficient. The estimated \bernstein polynomial $\widehat{\mathbb{B}f(t)}=\sum_{i=0}^m \hat f(i/m){m\choose i}t^i(1-t)^{m-i}$ aims at $\mathbb{B}f(t)$.
It is known that the best convergence rate of $\mathbb{B}f(t)$
to $f(t)$ is {\em at most} ${\cal O}(m^{-1})$ if $f$ has continuous second or even higher order derivatives on [0,1]. \cite{Buckland-1992-jrssc} proposed a density estimation with polynomials using grouped and ungrouped data with the help of some specified parametric models.

 Thanks to a result of \cite{Lorentz-1963-Math-Annalen}
there exists a \bernstein (type) polynomials $f_m(t; \bm p)\equiv  \sum_{i=0}^m p_{mi}\beta_{mi}(t)$, where $p_{mi}\ge 0$, $\beta_{mi}(t)=(m+1){m\choose i}t^i(1-t)^{m-i}$, $i=0,\ldots,m$, whose rate of convergence to $f(t)$ is {\em at least} ${\cal O}(m^{-r/2})$ if $f$ has a continuous $r$-th derivative
on [0,1] and $r\ge 2$. This is called a polynomial with ``positive coefficients'' in the literature of polynomial approximation.
\cite{Guan-arXiv-2014} introduced  the \bernstein polynomial model $f_m(t; \bm p)$ as a globally valid approximate parametric model of any underlying continuous density function with support $[0,1]$ and proposed a change-point method for selecting an optimal degree $m$. It has been shown that the rate of convergence to zero for the mean integrated squared error(MISE) of the maximum likelihood estimate of the density could be nearly parametric, ${\cal O}(n^{-1+\epsilon})$, for all $\epsilon>0$.
This method does not suffer from the boundary effect.

If the support of $f$ is different from [0,1] or even infinite, then we can choose an appropriate (truncation) interval $[a,b]$ so that $%F(b)-F(a)
\int_a^b f(x)dx\approx 1$
%$f$ is ignorable outside of $[a,b]$
\citep[see][]{Guan-arXiv-2014}. Therefore, we can treat $[a, b]$ as the support of $f$ and we can use the linearly transformed data $y_i=(x_i-a)/(b-a)$ in $[0, 1]$ to obtain estimate  $\hat g$ of  the PDF $g$ of $y_i$'s, respectively.
  Then %$\hat F(x)=\hat G\{(x-a)/(b-a)\}$ and
we estimate $f$ by $\hat f(x)=\hat g\{(x-a)/(b-a)\}/(b-a)$. In this paper, we will assume that the density $f$ has support $[0,1]$.

This \bernstein polynomial model $f_m(t; \bm p)$  is a finite mixture of the beta densities  $\beta_{mi}(t)$  of beta$(i+1, m-i+1)$, $i=0,\ldots,m$, with mixture proportions $\bm p=(p_{m0},\ldots, p_{mm})$. It has been shown that the \bernstein polynomial model can be used to fit a ungrouped dataset and  has the advantages of smoothness, robustness, and efficiency over the traditional methods such as the empirical distribution and the kernel density estimate \citep{Guan-arXiv-2014}.  Because these beta densities and their integrals are specified and free of unknown parameters, this structure of $f_m(t;\bm p)$ is convenient. It allows the grouped data to be approximately modeled by a mixture of $m+1$ {\em specific} discrete distributions. So the infinite dimensional ``parameter'' $f$ is approximately described by a finite dimensional parameter $\bm p$.  This and the nonparametric likelihood are similar in the sense that the underlying distribution function is approximated by a step function with jumps as parameters at the observations.

Due to the closeness of $f_m(t;\bm p)$ to $f(t)$, by the {\em acceptance-rejection} argument for generating pseudorandom numbers, almost all the observations in a sample from
$f(t)$ can be used as if they were from $f_m(t;\bm p)$. It will be shown in this paper that the maximizer of the likelihood based on the approximate model $f_m(t;\bm p)$ targets $\bm p_0$ which makes $f_m(t;\bm p_0)$ the unique best approximation of $f$. This acceptance-rejection  argument can be used to prove other asymptotic results
under an approximate model assumption.

In this paper  we shall study the asymptotic properties of the \bernstein polynomial density estimate based on grouped data and  ungrouped
raw data as a special case of grouping. A stronger result than that of \cite{Guan-arXiv-2014} about the rate of convergence of the proposed density estimate based on ungrouped raw data will be proved using a different argument. We shall also
compare the proposed estimate with those existing methods such as the kernel density, parametric MLE, and the MHDE via simulation study.

The paper is organized as follows. The \bernstein polynomial model for grouped data is introduced and is proved to be nested in Section \ref{sect: One-sample model}.
The EM algorithm for finding the approximate maximum likelihood estimates of the mixture proportions is derived in this section.
 Some asymptotic results
about the convergence rate of the proposed density estimate are given in Section \ref{sect: asymptotic results}.
%In Section \ref{sect: parameter estimate and choose optimal m} we present the estimation of some population parameters based on the proposed density estimate.
The methods for determining a lower bound for the model degree $m$ based on estimated mean and variance and for choosing the optimal degree $m$ are described in Section \ref{sect: parameter estimate and choose optimal m}. %this section.
In Section \ref{sect: simulations and example}, the proposed methods are compared with some
existing competitors through Monte Carlo experiments, and illustrated by the Chicken Embryo Data. The proofs of the theorems are relegated  to the Appendix.
\section{Likelihood for grouped data and EM algorithm}
\label{sect: One-sample model}
\subsection{The \bernstein polynomial model}
Let $C^{(r)}[0,1]$ be the class of functions which have $r$-th continuous derivative $f^{(r)}$ on $[0,1]$.
Like the normal model being backed up by the central limit theorem, the \bernstein polynomial model is supported by the following mathematical result which is a consequence of Theorem 1 of \cite{Lorentz-1963-Math-Annalen}. We denote the $m$-simplex by
$$\mathbb{S}_m=\Big\{\bm p=(p_{m0}, \ldots, p_{mm})^\tr\,:\,  p_{mj}\ge 0,\;\;  \sum_{j=0}^mp_{mj}= 1\Big\}.$$
\begin{theorem}\label{thm: approx of poly w pos coeff}
 If $f\in C^{(r)}[0,1]$, $\int_0^1 f(t)dt=1$, and $f(t)\ge \delta > 0$, then there exists a
 sequence of \bernstein type polynomials $f_m(t; \bm p)=\sum_{i=0}^m p_{mi}\beta_{mi}(t)$ with  $\bm p\in \mathbb{S}_m$, such that
\begin{equation}\label{eq: approx of f in Cr}
    |f(t)-f_m(t)|\le C(r,\delta,f) \Delta_m^{r}(t),\quad 0\le t\le 1,
\end{equation}
where $\Delta_m(t)=\max\{m^{-1}, \sqrt{{t(1-t)}/{m}}\}$ and the constant $C(r,\delta,f)$ depends on $r$, $\delta$, $\max_{t}|f(t)|$, and $\max_{t}|f^{(i)}(t)|$, $i=2,\ldots,r$, only. \end{theorem}
The uniqueness of the best approximation was proved by \cite{Passow-1977-JAT}.
%Note that $\beta_{mi}(t)=(m+1){m\choose i}t^i(1-t)^{m-i}$ is the density of the beta distributions, beta$(i+1, m-i+1)$, $i=0,\ldots,m$.
%It is easy to see that if $f$ is a {\em probability density function} on $[0,1]$, then such a \bernstein type polynomial $f_m(t)$ exists with $p_{mi}$'s satisfying %$\sum_{i=0}^m p_{mi}=1$.
%
Let $f$ be the density of the underlying distribution with support $[0,1]$.
We approximate $f$ using the Bernstein polynomial $f_m(t; \bm p)=\sum_{j=0}^mp_{mj}\beta_{mj}(t)$, where
$\bm p\in \mathbb{S}_m$.

Define ${\cal D}_m=\big\{f_m(t; \bm p)=\sum_{j=0}^mp_{mj}\beta_{mj}(t)\,:\, \bm p\in \mathbb{S}_m\big\}$.
\cite{Guan-arXiv-2014} showed that, for all $r\ge1$, ${\cal D}_m\subset {\cal D}_{m+r}$. So the {\em \bernstein polynomial model} $f_m(t; \bm p)$ of degree $m$ is
nested in all  \bernstein polynomial models of larger degrees.

Let $[0,1]$ be partitioned by $N$ class intervals
$\{(t_{i-1},t_i]\,:\, i=1,\ldots,N\}$, where
 $0=t_0<t_1<\dots<t_N=1$. The probability that a random observation falls in the $i$-th interval is approximately
\begin{equation}\label{eq: bernstein pol model for grouped data}
    \theta_{mi}(\bm p)=\int^{t_i}_{t_{i-1}}f_m(t;\bm p)dt=\sum_{j=0}^ma_{ij}p_{mj},
\end{equation}
where $a_{ij}=\mathcal{B}_{mj}(t_i)-\mathcal{B}_{mj}(t_{i-1})$,\; $i=1,\ldots,N$,\; $\mathcal{B}_{mj}(t)$ is the cumulative distribution function (CDF) of beta($j+1,m-j+1$),  $j=0,1,\ldots,m$,
and
$$\sum_{i=1}^N \theta_{mi}(\bm p)=\sum_{j=0}^mp_{mj}=1.$$
So the probability $\theta_{mi}(\bm p)$ is a mixture of a specific components $\{a_{i0},\ldots,a_{im}\}$ with unknown proportions $\bm p=(p_{m0},\ldots,p_{mm})$.

By Theorem 2$\cdot$1 of \cite{Guan-arXiv-2014}, the above \bernstein polynomial model (\ref{eq: bernstein pol model for grouped data}) of degree $m$ for grouped data is nested in a model of degree $m+r$, i.e.,  for all $r\ge 1$,
$$\theta_{mi}=\int^{t_i}_{t_{i-1}}\sum_{j=0}^m p_{mj} \beta_{mj}(t)dt=\int^{t_i}_{t_{i-1}}\sum_{j=0}^{m+r} p_{m+r,j} \beta_{m+r,j}(t)dt=\theta_{m+r,i},\quad i=1,\ldots,N.$$

\subsection{The \bernstein likelihood for grouped data}
In many applications, we only have the grouped data
$\{n_i, (t_{i-1},t_i]: i=1,\ldots,N\}$ available, where
 $0=t_0<t_1<\dots<t_N=1$ and $n_i=\#\{j\in (1,\ldots,n): x_j\in (t_{i-1},t_i]\}$, $i=1,\ldots,N$, and
  $x_1,\ldots,x_n$ is a random sample from a population having continuous density $f(x)$ on $[0,1]$. Our goal is to estimate
the unknown PDF $f$.
The loglikelihood of $(n_1,\ldots,n_N)$ is approximately
%\begin{equation}\label{eq: bernstein lik for grouped data}
%\mathscr{L}_G(\bm p)=\prod_{i=1}^N\Big\{\int^{t_i}_{t_{i-1}}f_m(t;\bm p)dt \Big\}^{n_i}=\prod_{i=1}^N %\Big[\sum_{j=0}^mp_{mj}\{\mathcal{B}_{mj}(t_i)-\mathcal{B}_{mj}(t_{i-1})\}\Big]^{n_i}.
%\end{equation}
%The loglikelihood for the grouped data is
\begin{equation}\label{eq: bernstein loglik for grouped data}
\ell_G(\bm p)=\sum_{i=1}^N n_i\log \Big[\sum_{j=0}^mp_{mj}\{\mathcal{B}_{mj}(t_i)-\mathcal{B}_{mj}(t_{i-1})\}\Big],
\end{equation}
where the mixture proportions $\bm p=(p_{m0},\ldots, p_{mm})$ are subject to the feasibility constraints
$\bm p\in \mathbb{S}_m$.
%\begin{equation}\label{constr: feasibility}
%p_{mj}\ge 0,\quad j=0,\ldots,m,\quad \sum_{j=0}^m p_{mj}=1.
%\end{equation}
For the ungrouped raw data $x_1,\ldots,x_n$, the loglikelihood is
\begin{equation}\label{eq: bernstein loglik for ungrouped data}
\ell_R(\bm p)=\sum_{i=1}^n  \log \Big\{\sum_{j=0}^mp_{mj}\beta_{mj}(x_i)\Big\}.
\end{equation}
If  we take the rounding error into account when the observations are rounded to the nearest value using the round half up  tie-breaking rule, then
\begin{equation}\label{eq: bernstein loglik for rounded data}
\ell_G(\bm p)=\sum_{i=-\infty}^\infty n_i\log \Big[\sum_{j=0}^mp_{mj}\{\mathcal{B}_{mj}(t_i)-\mathcal{B}_{mj}(t_{i-1})\}\Big],
\end{equation}
where $t_i=(i+1/2)/K$, $i=0,\pm 1,\pm 2,\ldots$, and $K$ is a positive integer such that any observation is rounded to $i/K$ for some integer $i$.

%where the mixture proportions $p_{m0},\ldots, p_{mm}$ are subject to (\ref{constr: feasibility}).
We shall call the maximizers $\tilde{\bm p}_G$ and $\hat{\bm p}_R$   of $\ell_G(\bm p)$  and $\ell_R(\bm p)$  the {\em maximum \bernstein likelihood estimates} (MBLE's) of $\bm p$ based on grouped and raw data,  respectively, and call
$\tilde f_B(t)=f_m(t; \tilde{\bm p}_G)$ and $\hat f_B(t)=f_m(t; \hat{\bm p}_R)$ the MBLE's of $f$ based on grouped and raw data,  respectively.

%\subsection{Quasi-Newton Method}
It should also be noted that
%, for equal-width classes with $\Delta t=t_{i+1}-t_i=1/N$,
as $N\to\infty$ and $\max  \{\Delta t_i\equiv t_{i}-t_{i-1}\,:\, i=1,\ldots,N\}\to 0$
the above loglikelihood (\ref{eq: bernstein loglik for grouped data}) reduces to the loglikelihood (\ref{eq: bernstein loglik for ungrouped data}) for ungrouped raw data. Specifically, $\lim_{\max \Delta t_i\to 0}\{\ell_G(\bm p)-\sum_{i=1}^N n_i\log\Delta t_i\} = \ell_R(\bm p)$.

If the underlying PDF $f$ is approximately $f_m(t; \bm p)=\sum_{i=0}^m p_{mi}\beta_{mi}(t)$ for some $m\ge 0$, then the distribution of the grouped data $(n_1,\ldots,n_N)$  is approximately multinomial
 with probability mass function
$$P(W_1=n_1,\ldots,W_N=n_N)= {n\choose n_1,\ldots,n_N}\prod_{i=1}^N\theta_{mi}^{n_i}(\bm p).$$
The MLE's of $\theta_{mi}$'s are
 $\hat\theta_{mi}=\frac{n_i}{n}$, $i=1,\ldots,N.$
So the MLE's $\hat p_{mj}$ of $p_{mj}$ satisfy the equations
$\sum_{j=0}^ma_{ij}\hat p_{mj}=\frac{n_i}{n}$, $i=1,\ldots,N$, and $(\hat p_{m0},\ldots,\hat p_{mm})\in \mathbb{S}_m$.
%where $a_{ij}=\mathcal{B}_{mj}(t_i)-\mathcal{B}_{mj}(t_{i-1})$, $i=1,\ldots,N$, $j=0,1,\ldots,m$.
Because $\hat p_{m0}=1-\sum_{j=1}^m\hat p_{mj}$,
$\hat p_{mj}$ satisfy equatins
$$\sum_{j=1}^m(a_{ij}-a_{i0})\hat p_{mj}=\frac{n_i}{n}-a_{i0},\quad i=1,\ldots,N,$$
and inequality constraints $\hat p_{mj}\ge 0$, $j=1,\ldots,m$, and $\sum_{j=1}^m\hat p_{mj}\le 1$. It seems not easy to algebraically solve the above system of equations with inequality constraints.
In the next section, we shall use an EM-algorithm to find the MLE of $\bm p$.
\subsection{The EM Algorithm}
Let $\delta_{ij}=1$ or 0 according to whether or not $x_i$ was  from beta$(j+1,m-j+1)$, $i=1,\ldots,n$,  $j=0,\ldots,m$.
We denote by $\bm z_i=(z_{i1},\ldots,z_{iN})^\tr$ the vector of indicators $z_{ij}=I\{x_i\in (t_{j-1},t_j]\}$, $i=1,\ldots,n$,  $j=1,\ldots,N$.
Then the expected value of $\delta_{ij}$ given $\bm z_i$ is
%$\E_{\bm p}(\delta_{ij})=p_j$, $\E(z_{ik}|\delta_{ij}=1)=\{\mathcal{B}_{mj}(t_k)-\mathcal{B}_{mj}(t_{k-1})\}$
%$$r_{jk}(\bm p)=\E_{\bm p}(\delta_{ij} z_{ik}|x)=\frac{p_j\{\mathcal{B}_{mj}(t_k)-\mathcal{B}_{mj}(t_{k-1})\}}{\sum_{h=0}^m %p_h\{\mathcal{B}_{mh}(t_k)-\mathcal{B}_{mh}(t_{k-1})\}}$$
$$r_{j}(\bm p,\bm z_i)\equiv\E_{\bm p}(\delta_{ij} |\bm z_i)=\frac{p_{mj}\prod_{l=1}^N\{\mathcal{B}_{mj}(t_l)-\mathcal{B}_{mj}(t_{l-1})\}^{z_{il}}}{\sum_{h=0}^m p_{mh}\prod_{l=1}^N\{\mathcal{B}_{mh}(t_l)-\mathcal{B}_{mh}(t_{l-1})\}^{z_{il}}}.$$
%$$\E_{\bm p}(\bm z_{ik}|\delta_{ij}=1, x)=\mathcal{B}_{mj}(t_k)-\mathcal{B}_{mj}(t_{k-1})$$
Note that $\sum_{j=0}^m\delta_{ij}=1$,
and the observations are $n_l=\sum_{i=1}^n\sum_{j=0}^m\delta_{ij}z_{il}=\sum_{i=1}^nz_{il}$, $l=1,\ldots,N$.
The likelihood of  $\delta_{ij}$  and $\bm z_i$  is
%$$\mathscr{L}_c=\prod_{i=1}^n\prod_{j=0}^m\prod_{l=1}^N [p_j\{\mathcal{B}_{mj}(t_k)-\mathcal{B}_{mj}(t_{k-1})\}]^{\delta_{ij}z_{ik}}$$
$$\mathscr{L}_c(\bm p)=\prod_{i=1}^n\prod_{j=0}^m \Big[p_{mj}\prod_{l=1}^N\{\mathcal{B}_{mj}(t_l)-\mathcal{B}_{mj}(t_{l-1})\}^{z_{il}}\Big]^{\delta_{ij}}.$$
The loglikelihood is then
%$$\ell_c(\bm p)=\sum_{i=1}^n\sum_{j=0}^m\sum_{k=1}^N  {\delta_{ij}z_{ik}}\log[p_j \{\mathcal{B}_{mj}(t_k)-\mathcal{B}_{mj}(t_{k-1})\}]$$
%$$\ell_c(\bm p)=\sum_{i=1}^n\sum_{j=0}^m\delta_{ij}   \log\left[p_j\prod_{k=1}^N %\{\mathcal{B}_{mj}(t_k)-\mathcal{B}_{mj}(t_{k-1})\}^{z_{ik}}\right]$$
$$\ell_c(\bm p)=\sum_{i=1}^n\sum_{j=0}^m\delta_{ij} \Big[ \log p_{mj}+\sum_{l=1}^N {z_{il}}\log\{\mathcal{B}_{mj}(t_l)-\mathcal{B}_{mj}(t_{l-1})\}\Big].$$
\textbf{E-Step}
Given $\tilde{\bm p}^{(s)}$, we have
\begin{eqnarray*}
% \nonumber to remove numbering (before each equation)
 Q(\bm p, \tilde{\bm p}^{(s)}) &=& \E_{\tilde{\bm p}^{(s)}}\{\ell(\bm p)|\bm z\} \\
  &=& \sum_{i=1}^n\sum_{j=0}^m r_{j}(\tilde{\bm p}^{(s)},\bm z_i)\Big[ \log p_{mj}+\sum_{l=1}^N {z_{il}}\log\{\mathcal{B}_{mj}(t_l)-\mathcal{B}_{mj}(t_{l-1})\}\Big].
\end{eqnarray*}
\textbf{M-Step}
Maximizing $Q(\bm p,\tilde{\bm p}^{(s)})$ with respect to $\bm p$ subject to constraint  $\bm p\in \mathbb{S}_m$ we have,
for $s=0,1,\ldots$,
%$$ p^{(s+1)}_j=\frac{1}{n}\sum_{i=1}^n{r_{j}(\bm p^{(s)},\bm %z_i)}=\frac{1}{n}\sum_{i=1}^n\frac{p_j^{(s)}\prod_{k=1}^N\{\mathcal{B}_{mj}(t_k)-\mathcal{B}_{mj}(t_{k-1})\}^{z_{ik}}}{\sum_{h=0}^m %p_h^{(s)}\prod_{k=1}^N\{\mathcal{B}_{mh}(t_k)-\mathcal{B}_{mh}(t_{k-1})\}^{z_{ik}}}.$$
%Alternatively, let $k_i$ be the integer such that $z_{ik_i}=I\{x_i\in (t_{k_i-1}, t_{ki}]\}=1$.
\begin{equation}\label{eq: EM iteration for p}
\tilde p^{(s+1)}_{mj}=\frac{1}{n}\sum_{i=1}^n{r_{j}(\tilde{\bm p}^{(s)},\bm z_i)}
%=\frac{1}{n}\sum_{i=1}^n\frac{p_j^{(s)}\{\mathcal{B}_{mj}(t_{k_i})-\mathcal{B}_{mj}(t_{{k_i}-1})\}}{\sum_{h=0}^m %p_h^{(s)}\{\mathcal{B}_{mh}(t_{k_i})-\mathcal{B}_{mh}(t_{{k_i}-1})\}}$$
%$$
=\frac{1}{n}\sum_{l=1}^N\frac{n_l\tilde p_{mj}^{(s)}\{\mathcal{B}_{mj}(t_{l})-\mathcal{B}_{mj}(t_{{l}-1})\}}{\sum_{h=0}^m \tilde p_{mh}^{(s)}\{\mathcal{B}_{mh}(t_{l})-\mathcal{B}_{mh}(t_{{l}-1})\}}.
\end{equation}
Starting with initial values $\tilde p_{mj}^{(0)}$, $j=0,\ldots,m$, we can use this iterative formula to obtain the maximum \bernstein likelihood
estimate $\tilde{\bm p}_G$. If
 the ungrouped raw data $x_1,\ldots,x_n$ are available, then the iteration \citep{Guan-arXiv-2014} is reduced to
\begin{equation}\label{eq: EM iteration for p ungrouped}
\hat p^{(s+1)}_{mj}
=\frac{1}{n}\sum_{i=1}^n\frac{\hat p_{mj}^{(s)}\beta_{mj}(x_{i})}{\sum_{h=0}^m \hat p_{mh}^{(s)}\beta_{mh}(x_{i})},\quad j=0,\ldots,m;\quad s=0,1,\ldots.
\end{equation}
%The convergence of $p^{(s+1)}_{mj}$'s to the maximizer  of $\ell(\bm p)$ will be proved by Theorem \ref{thm: convergence of EM algorithm} in Section \ref{sect: asymptotic results}.
\begin{comment}
If $p_{mj}^{(0)}=1/(m+1)$, $j=0,\ldots,m$, then
$$p^{(1)}_{mj}
=\frac{1}{n}\sum_{i=1}^n\frac{p_{mj}^{(0)}\beta_{mj}(x_{i})}{\sum_{h=0}^m p_{mh}^{(0)}\beta_{mh}(x_{i})}=p_{mj}^{(0)}\frac{1}{n}\sum_{i=1}^n \beta_{mj}(x_{i}),\quad j=0,\ldots,m;\quad s=0,1,\ldots.
$$
$$\frac{1}{n}\sum_{i=1}^n\frac{\beta_{mj}(x_{i})}{\sum_{h=0}^m p_{mh}^{(0)}\beta_{mh}(x_{i})}
=\frac{m+1}{n}\sum_{i=1}^n {m\choose j}x_i^j(1-x_i)^{m-j},\quad j=0,\ldots,m;\quad s=0,1,\ldots.
$$
$$\beta_{mj}(t)=(m+1){m\choose j}t^j(1-t)^{m-j},\quad \beta'_{mj}(t)=(j-mt)(m+1){m\choose j}t^{j-1}(1-t)^{m-j-1}$$
$$\beta''_{mj}(t)=(j-mt)\{j-1-(m-2)t\}(m+1){m\choose j}t^{j-2}(1-t)^{m-j-2}-m(m+1){m\choose j}t^{j-1}(1-t)^{m-j-1}$$
$$\beta''_{mj}(t)=\left[(j-mt)\{j-1-(m-2)t\}-mt(1-t)\right](m+1){m\choose j}t^{j-2}(1-t)^{m-j-2}$$
$$\beta''_{mj}(t)=\{j(j-1)-2j(m-1)t+m(m-1)t^2\}(m+1){m\choose j}t^{j-2}(1-t)^{m-j-2}$$
\end{comment}
%\subsection{Convergence of $p_{mj}^{(s)}$}
The following theorem shows the convergence of the EM algorithm and is proved in the Appendix.
\begin{theorem}\label{thm: convergence of EM algorithm}
(i) Assume $\hat p^{(0)}_{mj}>0$, $j=0,1,\ldots,m$, and  $\sum_{j=0}^m \hat p^{(0)}_{mj}=1$.
Then as $s\to\infty$, $\hat {\bm p}^{(s)}$ converges to the unique maximizer $\hat{\bm p}_R$ of $\ell_R(\bm p)$.
(ii) Assume $\tilde p^{(0)}_{mj}>0$, $j=0,1,\ldots,m$, and  $\sum_{j=0}^m \tilde p^{(0)}_{mj}=1$.
Then as $s\to\infty$, $\tilde {\bm p}^{(s)}$ converges to the unique maximizer $\tilde{\bm p}_G$ of $\ell_G(\bm p)$.
\end{theorem}

\section{%Asymptotic Results
Rate of Convergence of the Density Estimate}\label{sect: asymptotic results}
%\subsection{Strong Consistency of $\hat{\bm p}$}
% \subsection{Weak Convergence of Stochastic Processes Associated with $\hat f_B$ and $\hat F_B$}
In this section we shall
%prove an improved version of Theorem 4.1 of \cite{Guan-arXiv-2014} and a generalization to the grouped data
state results about the convergence rate of the density estimates which will be proved in the Appendix. Unlike most asymptotic results about maximum likelihood method which assume exact parametric models, we will
show our results under the approximate model $f_m(t;\bm p)=\sum_{j=0}^mp_{mj}\beta_{mj}(t)$. %=\beta_{m0}(t)+\bm p_m^\tr \bar{\bm\beta}_m(t)$
For a given $\bm p_0$, we define the  norm
$$\Vert \bm p\Vert^2_B=\int\frac{\{f_m(t; \bm p)\}^2}{f_m(t; \bm p_0)}dt.
%=\bm p^\tr B_m(\bm p_0) \bm p,
$$
%where $B_m(\bm p_0)=\{b_{ij}(\bm p_0)\}$ is an $(m+1)\times(m+1)$ positive definite matrix with entries
%$$b_{ij}(\bm p_0)=
%\int_0^1\frac{\beta_{mi}(t)\beta_{mj}(t)}{f_m(t; \bm p_0)}dt=\frac{(m+1)^2{m\choose i}{m\choose j}}{(2m+1){2m\choose i+j}}
%\int_0^1\frac{\beta_{2m,i+j}(t)}{f_m(t; \bm p_0)}dt ,\quad  0\le i,j\le m.$$
\begin{comment}
$$b_{ij}(\bm p_0)=
\int_0^1\frac{\beta_{mi}(t)\beta_{mj}(t)}{f_m(t; \bm p_0)}dt ,\quad  0\le i,j\le m.$$
So we have $B_m(\bm p_0)\bm p_0=\bm 1$.
Define $\bm\beta_m(t)=(\beta_{m0}(t),\ldots,\beta_{mm}(t))^\tr$, $\tilde{\bm\beta}_m(t)=\bm\beta_m(t)/\sqrt{f_m(t;\bm p_0)}$,
and $${\cal B}_m(t)=\tilde{\bm\beta}_m(t)\tilde{\bm\beta}_m^\tr (t)=\frac{{\bm\beta}_m(t){\bm\beta}_m^\tr (t)}{f_m(t; \bm p_0)}$$ Then
$$B_m(\bm p_0)=\int_0^1 {\cal B}_m(t)dt.$$
\end{comment}
%Note that $f_m(t; \bm p)-f_m(t; \bm p')=f_m(t; \bm p-\bm p')$.
The squared distance between $\bm p$ and $\bm p_0$ with respect to norm $\Vert \cdot\Vert_B$ is
$$\Vert \bm p-\bm p_0\Vert^2_B=\int\frac{\{f_m(t; \bm p)-f_m(t; \bm p_0)\}^2}{f_m(t; \bm p_0)}dt.
%=(\bm p-\bm p_0)^\tr B_m(\bm p_0) (\bm p-\bm p_0).
$$

With the aid of the acceptance-rejection argument for generating pseudorandom numbers in the Monte Carlo method we have the following lemma which may be of independent interest.
\begin{lemma} \label{lemma: acceptance-rejection argument}
    Let  $f\in C^{(2k)}[0,1]$ for some positive integer $k$,   $f(t)\ge \delta > 0$, and $f_m(t)=f_m(t; \bm p_0)$ be the unique best approximation of degree $m$ for $f$.
Then a sample $x_1,\ldots,x_n$ from $f$ can be arranged so that the first $\nu_m$ observations can be treated as if they were from $f_m$. Moreover, for all $\bm p$ such that
%$\Vert \bm p-\bm p_0\Vert^2_B\le C {(\log n)^2}/{n}$ for some $C>0$, and
$f_m(x_j;\bm p)\ge \delta'>0$, $j=1,\ldots,n$,
\begin{equation}\label{eq: approx of likelihood}
    \ell_R(\bm p)=\sum_{i=1}^n\log f_m(x_i; \bm p)=\tilde\ell_R(\bm p)+R_{mn},
\end{equation}
where $\tilde\ell_R(\bm p)=\sum_{i=1}^{\nu_m}\log f_m(x_i; \bm p)$, and
\begin{eqnarray}\label{eq: est of nu_m}
% \nonumber to remove numbering (before each equation)
  \nu_m &=&  n-{\cal O}(nm^{-k})-{\cal O}\left(\sqrt{nm^{-k}\log\log n}\right),\quad a.s.,\\
  \label{eq: est of R(mn)}
  |R_{mn}| &=&  {\cal O}(nm^{-k})+{\cal O}\left(\sqrt{nm^{-k}\log\log n}\right),\quad a.s..
\end{eqnarray}
\end{lemma}
\begin{remark}
So $\tilde\ell_R(\bm p)$ is an ``exact'' likelihood of $x_1,\ldots,x_{\nu_m}$ while $\ell_R(\bm p)=\sum_{i=1}^n\log f_m(x_i; \bm p)$ is an approximate likelihood of the complete data $x_1,\ldots,x_n$ which can be viewed as a slightly  contaminated sample from $f_m$. Maximizer $\hat{\bm p}$ of $\ell_R(\bm p)$  approximately  maximizes
$\tilde\ell_R(\bm p)$. Hence $f_m(t;\hat{\bm p})$ targets at $f_m(t;\bm p_0)$ which is a best approximate of $f$.
\end{remark}

For density estimation based on the  raw data we have the following result.
\begin{theorem}\label{thm: convergence rate for raw data}
%For each $\epsilon>0$, $r_n=n^{(\epsilon-1)/2}$,
Suppose that the PDF $f\in C^{(2k)}[0,1]$ for some positive integer $k$,   $f(t)\ge \delta > 0$, and $m={\cal O}(n^{1/k})$.   As $n \to\infty $, with probability one  the  maximum
value of $\ell_R(\bm p)$ is attained by some $\hat{\bm p}_R$ in the interior of
$\mathbb{B}_m(r_n)=\{\bm p\in \mathbb{S}_m\,:\,\Vert \bm p-\bm p_0\Vert^2_B \le r_n^2\}$, %n^{-2r}\}$,
where $r_n=\log n/\sqrt{n}$ and
$\bm p_0$ makes $f_m(\cdot;\bm p_0)$ the unique best approximation of degree $m$. %( which is in the interior of $\Delta_m$? what if $p_{0i}=0$ for some $i$?).
\end{theorem}
\begin{comment}
\begin{rem}
From this theorem we have that there is a positive constant $C$ such that
$$\E\int\frac{\{f_m(t; \hat{\bm p}_R)-f_m(t; \bm p_0)\}^2}{f_m(t; \bm p_0)}dt \le C r_n^2=C\frac{(\log n)^2}{n}.$$
%This is an almost parametric rate of convergence.
Because $f$ is bounded  there is a positive constant $C$ such that $$\E \int\{f_m(t; \hat{\bm p}_R)-f_m(t; \bm p_0)\}^2dt \le C r_n^2=C\frac{(\log n)^2}{n}.$$
This is an almost parametric rate of convergence.
\end{rem}
\end{comment}
\begin{theorem}\label{cor: mise for raw data}
Suppose that the PDF $f\in C^{(2k)}[0,1]$ for some positive integer $k$,   $f(t)\ge \delta > 0$, and $0<c_0n^{1/k}\le m\le c_1 n^{1/k}<\infty$. Then there is a positive constant  $C$   such that
\begin{eqnarray}\label{eq: weighted mise for raw data}
\E\int\frac{\{f_m(t; \hat{\bm p}_R)-f(t)\}^2}{f(t)}dt
%&\le& \E\int\frac{\{f_m(t; \hat{\bm p}_R)-f_m(t; \bm p_0)\}^2}{f_m(t; \bm p_0)}\frac{f_m(t; \bm p_0)}{f(t)}dt\\
%&&+\E\int\frac{\{f_m(t; {\bm p}_0)-f(t)\}^2}{f(t)}dt\\
&\le& % C_1 r_n^2+C_2 m^{-k-\alpha/2}\le
 C\frac{(\log n)^2}{n}.
\end{eqnarray}
%This is an almost parametric rate of convergence.
Because $f$ is bounded  there is a positive constant $C$ such that
\begin{eqnarray}\label{eq: mise for raw data}
\mathrm{MISE}(\hat f_B)&=&\E \int\{f_m(t; \hat{\bm p}_R)-f(t)\}^2dt \le %C r_n^2=
C\frac{(\log n)^2}{n}.
\end{eqnarray}
\end{theorem}
Note that (\ref{eq: weighted mise for raw data}) is a stronger result than (\ref{eq: mise for raw data}) which is an almost parametric rate of convergence for MISE.
\cite{Guan-arXiv-2014} showed a similar result under another set of conditions. The best parametric rate is ${\cal O}(n^{-1})$ that can be attained by the parametric density estimate under some regularity conditions.

For $\theta_{mi}(\bm p)=\sum_{j=0}^mp_{mj}\{\mathcal{B}_{mj}(t_i)-\mathcal{B}_{mj}(t_{i-1})\}$, we define norm
$$\Vert \bm p\Vert^2_G=\sum_{i=1}^{N}\frac{\theta_{mi}^2(\bm p)}{\theta_{mi}(\bm p_0)}.$$ The squared distance between $\bm p$ and $\bm p_0$ with respect to norm
$\Vert\cdot\Vert_G$ is
$$\Vert \bm p-\bm p_0\Vert^2_G=\sum_{i=1}^{N}\frac{\{\theta_{mi}(\bm p)-\theta_{mi}(\bm p_0)\}^2}{\theta_{mi}(\bm p_0)}.$$
By the mean value theorem, we have
$$\mathcal{B}_{mj}(t_i)-\mathcal{B}_{mj}(t_{i-1})=\int_{t_{i-1}}^{t_i}\beta_{mj}(t)dt=
 \beta_{mj}(t_{mij}^*)\Delta t_i,\quad i=1,\ldots,N;\;j=0,\ldots,m,$$
%$$\theta_{mi}(\bm p)=\sum_{j=0}^m p_{mj}\int_{t_{i-1}}^{t_i}\beta_{mj}(t)dt=
%\sum_{j=0}^m p_{mj}\beta_{mj}(t_{mij}^*)\Delta t_i,$$
where $\Delta t_i=t_i-t_{i-1}$ and $t_{mij}^*\in [t_{i-1},t_i]$. Thus $\Vert\bm p- \bm p_0\Vert_G$ is a Riemann sum
%Assume that the class intervals have equal width $\Delta t_i=1/N$ and $t_i=i/N$, $i=0,1,\ldots,N$.
%$$\theta_{mi}(\bm p)-\theta_{mi}(\bm p_0)=\sum_{j=0}^m(p_{mj}- p_{mj}^{(0)}) \beta_{mj}(t_{mij}^*)\Delta t_i.$$
%$$\frac{\{\theta_{mi}(\bm p)-\theta_{mi}(\bm p_0)\}^2}{\theta_{mi}(\bm p_0)}
%=\frac{\{\sum_{j=0}^m(p_{mj}- p_{mj}^{(0)}) \beta_{mj}(t_{mij}^*)\Delta t_i\}^2}{\sum_{j=0}^m p_{mj}^{(0)}\beta_{mj}(t_{mij}^*)\Delta t_i}.$$
\begin{equation}\label{eq: Riemann sum expression}
   \Vert\bm p- \bm p_0\Vert_G^2= \sum_{i=1}^{N}\psi_m(t_{mij}^*)\Delta t_i\approx \int_0^1 \psi_m(t)dt=\Vert \bm p-\bm p_0\Vert^2_B,
\end{equation}
where
$$\psi_m(t)%=\frac{(\Delta p\beta)^2}{p_0\beta}
=\frac{\{\sum_{j=0}^m(p_{mj}- p_{mj}^{(0)}) \beta_{mj}(t)\}^2}{\sum_{j=0}^m p_{mj}^{(0)}\beta_{mj}(t) }.$$
For grouped data we have the following.
\begin{theorem}\label{thm: convergence rate for grouped data}
% For each $\epsilon>0$, $r_n=n^{(\epsilon-1)/2}$, then,
Suppose that the PDF $f\in C^{(2k)}[0,1]$ for some positive integer $k$,   $f(t)\ge \delta > 0$, and $m={\cal O}(n^{1/k})$.
  As $n \to\infty $, with probability one  the  maximum
value of $\ell_G(\theta_m)$ is attained at $\tilde{\bm p}_G$ in the interior of
$\mathbb{B}_m(r_n)=\{\bm p\in \mathbb{S}_m\,:\,\Vert \bm p-\bm p_0\Vert^2_G \le r_n^2\}$, %n^{-2r}\}$,
where $r_n=\log n/\sqrt{n}$  and
$\bm p_0$ makes $f_m(\cdot;\bm p_0)$ the unique best approximation. %( which is in the interior of $\Delta_m$? what if $p_{0i}=0$ for some $i$?).
\end{theorem}
For the relationship between the norms $\Vert\bm p- \bm p_0\Vert_B^2$ and $\Vert\bm p- \bm p_0\Vert_G^2$, we have the following result.
\begin{theorem}\label{thm: relationship between the two norms}
Suppose that the PDF $f\in C^{(2k)}[0,1]$ for some positive integer $k$, and  $f(t)\ge \delta > 0$.
Let $\bm p_0\in \mathbb{S}_m$ be the one that makes $f_m(\cdot;\bm p_0)$ the unique best approximation of $f$. Then for all $\bm p\in \mathbb{S}_m$, we have
$$\Vert \bm p-\bm p_0\Vert^2_B=\Vert\bm p- \bm p_0\Vert_G^2+{\cal O}(m^4 \max_i \Delta t_i^2).$$
\end{theorem}
%\begin{rem}
For a  grouped data based estimate $\tilde{\bm p}_G$,  the rate of convergence of   $\Vert\tilde{\bm p}_G- \bm p_0\Vert_G^2$ to zero is ${\cal O}((\log n)^2/n)$. However the rate of convergence of
$\Vert \tilde{\bm p}_G-\bm p_0\Vert^2_B$ to zero depends on that of $\max_i \Delta t_i$.
%If the number $N=N_n$ of class intervals can be arbitrarily large, i.e., $N_n$ can approach
%infinity as fast as it can so that $\Vert\hat{\bm p}_G- \bm p_0\Vert_B^2={\cal O}((\log n)^2/n)
%+{\cal O}(m^4\max \Delta t_i^2)$.
For equal-width classes, $\Delta t_i=1/N$,  and $N=n^{1/2+2/k}$, we have
$\Vert\tilde{\bm p}_G- \bm p_0\Vert_B^2= {\cal O}((\log n)^2/n)
+{\cal O}(m^4/n^{1+4/k})$. Thus
$\Vert\tilde{\bm p}_G- \bm p_0\Vert_B^2={\cal O}((\log n)^2/n)$ if
  $m={\cal O}(n^{1/k})$. %, $\Vert\hat{\bm p}_G- \bm p_0\Vert_B^2={\cal O}((\log n)^2/n)$ if $m={\cal O}(\log n)$.
If $k$ is large, then $N\approx\sqrt{n}$.
%This is an advantage over other methods such as the
%approximate minimum Hellinger distance estimate \cite{Lin-and-He-2006-bka} which is a parametric method.
%\end{rem}
\begin{theorem}\label{cor: mise for grouped data}
Suppose that the PDF $f\in C^{(2k)}[0,1]$ for some positive integer $k$,   $f(t)\ge \delta > 0$, and $0<c_0n^{1/k}\le m\le c_1 n^{1/k}<\infty$.
Then we have
\begin{eqnarray}\label{eq: weighted mise for grouped data}
\E\int\frac{\{f_m(t; \tilde{\bm p}_G)-f(t)\}^2}{f(t)}dt&= &
 {\cal O}((\log n)^2/n)  +{\cal O}(m^4 \max_i \Delta t_i^2).
\end{eqnarray}
%This is an almost parametric rate of convergence.
Also, because $f$ is bounded,
\begin{eqnarray}\nonumber
\mathrm{MISE}(\hat f_B)&=&\E \int\{f_m(t; \tilde{\bm p}_G)-f(t)\}^2dt \\
\label{eq: mise for grouped data}
&=& {\cal O}((\log n)^2/n)
+{\cal O}(m^4 \max_i \Delta t_i^2).
\end{eqnarray}
\end{theorem}
\section{Model Degree Selection}\label{sect: parameter estimate and choose optimal m}

\cite{Guan-arXiv-2014} showed that the model degree $m$ is bounded below approximately by $m_b=\max\left\{1,\ceiling{\mu(1-\mu) /\sigma^{2}-3}\right\}$.  Based on the grouped data,
the lower bound $m_b$ can be estimated by
$\tilde m_b =\max\left\{1,\ceiling{\tilde\mu(1-\tilde\mu) /\tilde\sigma^{2}-3}\right\}$, where
$$\tilde\mu=\frac{1}{n}\sum_{i=1}^N n_i t_i^*,\quad \tilde\sigma^2=\frac{1}{n-1}\sum_{i=1}^N n_i(t_i^*-\tilde\mu)^2=\frac{1}{n-1}\left(\sum_{i=1}^N n_it_i^{*2}-n\tilde\mu^2\right),$$
$$
t_i^*=(t_{i-1}+t_i)/2,\quad i=1,\ldots,N.$$

Due to overfitting the model degree $m$ cannot be arbitrarily large.
With the estimated $\tilde m_b$, we choose a proper set of nonnegative consecutive integers, $M=\{m_0,m_0+1,\ldots,m_0+k\}$ such that $m_0<\tilde m_b$.
Then we can estimate an optimal degree $m$ using the method of change-point estimation as proposed by \cite{Guan-arXiv-2014}. For each $m_i=m_0+i$ we use the EM algorithm to find the
 MBLE $\tilde{\bm p}_{m_i}$ and calculate $\ell_i=\ell(\tilde{\bm p}_{m_i})$. Let $y_i=\ell_{i}-\ell_{i-1}$, $i=1,\ldots,k$.
The $y_i$'s are nonnegative because the \bernstein polynomial models are nested.
%the $\hat m= m_{\hat\tau}$ with $\hat\tau$ being the estimated  change-point of $y_1,\ldots,y_k$.
\cite{Guan-arXiv-2014} suggested that  $y_1,\ldots,y_\tau$ be treated as  exponentials with mean $\mu_1$ and  $y_{\tau+1},\ldots,y_k$ be treated  as  exponentials with mean $\mu_0$, where $\mu_1>\mu_0$, so that $\tau$ is a change point and $m_\tau$ is the optimal degree and use the change-point detection method \citep[see  Section 1.4 of][]{Csorgo1997a} for exponential
model to find a change-point estimate $\hat \tau$. Then we estimate the optimal $m$ by $\hat m=m_{\hat \tau}$. Specifically, $\hat\tau=\arg\max_{1\le \tau\le k}\{R(\tau)\}$, where    the
likelihood ratio  of $\tau$ is
%$R(\tau)=-\tau\log\{S_\tau/\tau\}-(k-\tau)\log\{(S_k-S_\tau)/(k-\tau)\}+k\log\{S_k/k\}$, $\tau=1,\ldots,k,$ and
% $S_\tau=\sum_{j=1}^\tau y_j=\sum_{j=1}^\tau(\ell_j-\ell_{j-1})=\ell_\tau-\ell_0$, $\tau=1,\ldots,k$. It is obvious that
$$R(\tau)=k\log\left(\frac{\ell_k-\ell_0}{k}\right)-\tau\log\left(\frac{\ell_\tau-\ell_0}{\tau}\right)-(k-\tau)\log\left(\frac{\ell_k-\ell_\tau}{k-\tau}\right),
\quad\tau=1,\ldots,k.$$
If $R(\tau)$ has multiple maximizers, we choose the smallest one as $\hat\tau$.
\section{Simulation Study and Example}\label{sect: simulations and example}
\subsection{Simulation}
The distributions used for generating pseudorandom numbers and the parametric models used for density estimation are as following.
\begin{itemize}
    \item [(i)]  Uniform(0,1): the uniform distribution with $\mu=1/2$ and $\sigma^2=1/12$  as a special {beta} distribution beta(1,1). The parametric model is  the {beta} distribution beta($\alpha$, 1).
    \item [(ii)] Exp(1): the exponential distribution  with mean $\mu=1$ and variance $\sigma^2=1$. We truncate this distribution by the  interval $[a,b]=[0, 4]$.
The parametric model is the exponential distribution  with mean $\mu=\theta$.
    \item [(iii)] Pareto(4, 0.5): The Pareto distribution with shape parameter $\alpha=4$ and scale parameter $x_0=$ 0.5  which is treated as known parameter.
The mean and variance are, respectively, $\mu=\alpha x_0/(\alpha-1)=2/3$ and $\sigma^2=  x_0^2\alpha/[(\alpha-1)^2(\alpha-2)]=1/18$. We truncate this distribution by the  interval $[a,b]=[x_0, \mu+4\sigma]\approx$ [0.5, 1.6095]. The parametric model is Pareto($\alpha$, 0.5).
    \item [(iv)] NN($k$): the {\em nearly normal}
    %$\bar U_k[0,1]$:
      distribution  of $\bar u_k=(u_1+\cdots+u_{k})/k$ with $u_1,\ldots,u_{k}$ being independent uniform(0,1) random variables.
%The density of $\bar u_k$ is denoted by $\psi_k(t)$.
The lower bound is $m_b=%\mu(1-\mu)/\sigma^2-3=
      3(k-1)$. We used the {normal} distribution N($\mu$, $\sigma^2$) as the parametric model.
    \item [(v)]  N($0, 1$): the standard {normal} distribution  truncated   by the interval $[a,b]=[-4, 4]$. %$[\mu-k\sigma, \mu+k\sigma]$ with $k=5$.
%    Thus $m_b=23$.
The parametric model is N($\mu, \sigma^2$).
    \item [(vi)]  Logistic(0, 0.5): the logistic distribution with location $\mu=0$ and scale $s=$ 0.5 so  that $\sigma^2=(s\pi)^2/3=\pi^2/12$. We truncate this distribution by the  interval $[a,b]=[\mu-4\sigma, \mu+4\sigma]\approx$ [-2.9619, 2.9619]. The parametric model is Logistic($\mu$, $s$).
\end{itemize}

Except the normal distribution,  the above parametric models were chosen for the simulation because the  CDF's have
close-form expressions so that the expensive numerical integrations can be avoided for the MHDE and the MLE.

From each distribution we generated 500 samples of size $n=50, 100, 200$, and $500$ and the grouped data using $N=5, 10, 10$ and $20$ equal-width class intervals, respectively. The model degree $m$
were selected using the change-point method from $\{1,2,\ldots,40\}$.

From the results of \cite{Guan-arXiv-2014} we see that the \bernstein polynomial method is much better than the kernel density for ungrouped data.
  The AMHDE is approximation for MHDE.
So we only compare kernel, the MLE, the MHDE,  and the proposed MBLE. For the kernel density estimate
$\hat{f}_K(x) =
% \frac{1}{n}\sum_{i=1}^n K_h (x - x_i) \quad =
\frac{1}{nh} \sum_{i=1}^n K\Big(\frac{x-x_i}{h}\Big),$ we used  normal kernel $K(x)=e^{-x^2/2}/\sqrt{2\pi}$ and the commonly recommended  method of \cite{Sheather-and-Jones-1991-JRSSB} to choose the bandwidth $h$. Because
$\E[\hat{f}_K(x)] =
% \frac{1}{n}\sum_{i=1}^n K_h (x - x_i) \quad =
\frac{1}{h} \int_{-\infty}^\infty K\Big(\frac{x-y}{h}\Big)f(y)dy=(K_h*f)(x).$
    This is the  {\em convolution} of $f$ and the scaled kernel $K_h(\cdot)=K(\cdot/h)/h$. So no matter how the bandwidth $h$ is chosen, there is always trade-off
    between the bias and the variance.
\renewcommand{\baselinestretch}{1.0}
\begin{table}
  \centering
  \caption{Estimated mean and variance of $\hat m$, and mean integrated squared errors (MISE's) of the  kernel density $\hat f_K$, the  MLE  $\hat f_{\ML}$, the  MHDE $\hat f_{\MHD}$,    and the proposed maximum \bernstein likelihood estimate (MBLE) $\hat f_{\mathrm{B}}$   based 500 simulated samples of size $n$ which are grouped by $N$  equal-width class intervals, respectively.}\label{tbl1}
\begin{tabular}{lcccccc}
  \hline
  % after \\: \hline or \cline{col1-col2} \cline{col3-col4} ...
    &&&\multicolumn{4}{c}{MISE}\\ \cline{4-7}
\Big.&$\E(\hat m)$&$\Var(\hat m)$&$\hat f_{\mathrm{B}}$& $\hat f_{\mathrm{K}}$ & $\hat f_{\ML}$ & $\hat f_{\MHD}$ \\\hline
&\multicolumn{6}{c}{$n=50$, $N=5$}\\
Beta(1,1)&14.91&~3.95&0.3898&~2.5722&0.0193&0.0222\\
Exp(1)&14.56&~9.14&0.0447&~0.7502&0.0018&0.0034\\
Pareto&12.29&29.01&0.7855&19.6962&0.0392&0.0793\\
NN$(4)$&12.04&15.42&0.0556&~8.5549&0.0653&0.103\\
N$(0,1)$&14.25&11.10&0.0007&~0.1603&0.0008&0.0012\\
Logistic&13.79&19.27&0.0022&~0.2689&0.0014&0.0024\\
&\multicolumn{6}{c}{$n=100$, $N=10$}\\
Beta(1,1)&12.96&47.08&0.0972&~0.0558&0.0096&0.0118\\
Exp(1)&~~9.42&37.24&0.0091&~0.2377&0.0011&0.0027\\
Pareto&~~8.51&~~9.78&0.1009&18.6903&0.0222&0.0613\\
NN$(4)$&10.24&~~6.72&0.0217&~1.4357&0.0232&0.0411\\
N$(0,1)$&13.77&~~6.58&0.0004&~0.0357&0.0003&0.0005\\
Logistic&12.02&13.42&0.0012&~0.0992&0.0007&0.0012\\
&\multicolumn{6}{c}{$n=200$, $N=10$}\\%\hline
Beta(1,1)&15.88&48.93&0.0741&~~1.2338&0.0045&0.0051\\
Exp(1)&~~9.24&41.36&0.0068&~~0.4547&0.0007&0.0017\\
Pareto&~~8.63&~~9.85&0.0661&34.3823&0.0123&0.0323\\
NN$(4)$&10.11&~~3.92&0.0128&~~4.0956&0.0125&0.0213\\
N$(0,1)$&14.08&~~5.74&0.0003&~~0.0907&0.0002&0.0003\\
Logistic&13.15&14.07&0.0007&~~0.1936&0.0004&0.0006\\
&\multicolumn{6}{c}{$n=500$, $N=20$}\\%\hline
Beta(1,1)&10.18&49.84&0.0192&~~0.0226&0.0021&0.0024\\
Exp(1)&~~4.40&~~3.55&0.0006&~~0.2331&0.0005&0.0015\\
Pareto&~~8.67&~~1.94&0.0181&16.0924&0.0083&0.0253\\
NN$(4)$&~~9.97&~~2.75&0.0059&~~0.5994&0.0058&0.0110\\
N$(0,1)$&14.41&~~2.94&0.0001&~~0.0329&0.0001&0.0001\\
Logistic&13.26&~~5.15&0.0003&~~0.0905&0.0001&0.0003\\
  \hline
\end{tabular}
\end{table}
\renewcommand{\baselinestretch}{2}

Table \ref{tbl1} presents the simulation results of the density estimations. As expected, the proposed \bernstein polynomial method performs much better than the
kernel density method and is similar to the other two parametric methods. Table \ref{tbl1} also shows the estimated mean and variance of the optimal model degree selected by the change-point method. It seems that the performance of the estimated optimal model degree $\hat m$ is satisfactory.

It should be noted that the density $\psi_k(t)$ of NN($k$) satisfies $\psi_k(t)\in C^{(k-2)}[0,1]$ but $\psi_k(t)\notin C^{(k-1)}[0,1]$ for $k\ge 2$. In fact, when, $k\ge 2$,  $\psi_k(t)$ is a piecewise  polynomial function of degree $(k-1)$ defined on pieces $[i/k, (i+1)/k)$, $i=0,1,\ldots,k-1$. Except NN(k) all the other population densities  have continuous derivatives of all orders on their supports. In the simulation, we used the normal distributions as the parametric models of NN($4$). Here both the normal and the \bernstein polynomial are {\em approximate  models}. In fact,
in most applications the normal distribution is an approximate model due to the central limit theorem.  We did a simulation on the goodness-of-fit of the normal distribution to the sample from NN($4$). In this simulation, we generate $5,000$ samples of size $n$ from NN($4$). We ran the Kolmogorov-Smirnov test for each sample. For $n=50, 100, 200$, and $500$ the average of the $p$-values are, respectively, 0.7884, 0.7875, and 0.7470; and the numbers of $p$-values among the 5000 that are smaller than 0.05 are, respectively, 3, 2, 0, and 2.
% n=200, k=4, B=5000
%$ave.pval= 0.7874738; $reject.prob =  0
%min(res$pval) = 0.05544591
% n=500, k=4, B=5000
%$ave.pval=0.7469621, $reject.prob= 4e-04 =2/5000,
So the normal distribution will accepted as the parametric model for NN(4) almost all the time.
The performance of the proposed MBLE for samples from NN(4) is even better than that of the MLE when sample size $n$ is small.
 \subsection{The Chicken Embryo Data}
The  chicken embryo
data contain the  {number} of hatched eggs on  {each day} during the 21 days of incubation
period. The  {times of hatching} ($n=43$) are treated as  {grouped}
by intervals  with {equal width of one day}.  The  data were studied first by \cite{Jassim-et-al-1996}. %Jassim et al. (1996).
 \cite{Kuurman-et-al-2003} % Kuurman et al. (2003)
  and
 \cite{Lin-and-He-2006-bka} %Lin and He (2006)
 also analyzed the data using the MHDE, in addition to other methods assuming some parametric mixture models including Weibull model. The latter used the AMHDE to fit the data by Weibull mixture model. The estimated density using the proposed method is close to the parametric
MLE.

Applying the proposed method of this paper, we truncated the distribution using $[a,b]=[0, 21]$ and selected the optimal model degree $\hat m=13$ from $\{2, 3,\ldots, 50\}$ using the change-point method.
\begin{figure}
\begin{center}
  % Requires \usepackage{graphicx}
%  \makebox{
\includegraphics[width=5.0in]{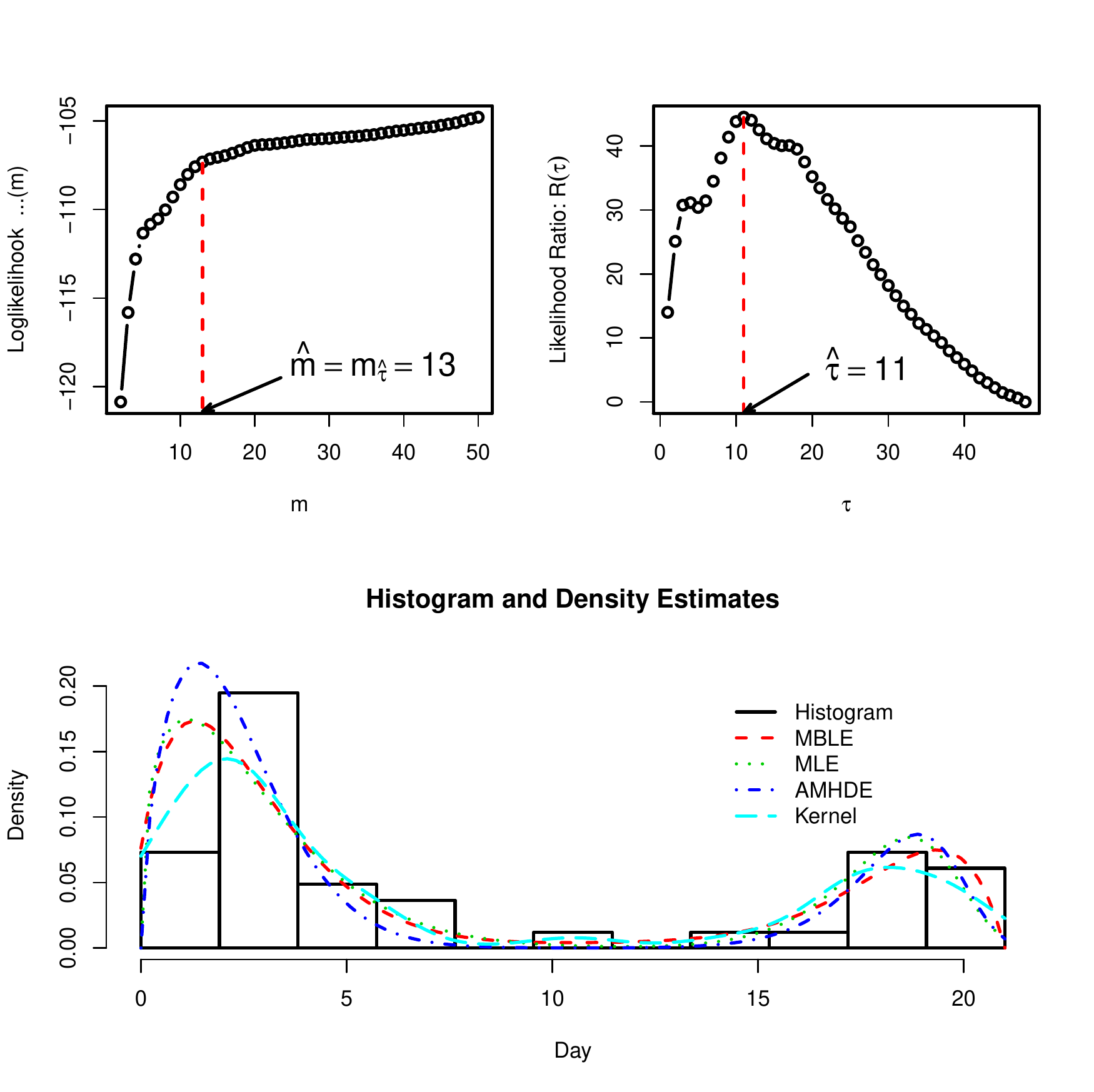}%}
\caption{\label{fig: chicken-embryo-data-densities} Upper panel left: the loglikelihood $\ell(m)$; Upper panel right: the likelihood ratio $R(\tau)$ for change-point
of the increments of the loglikelihoods $\ell(m)$; Lower panel: the density estimates for the chicken embryo data.
}
\end{center}
\end{figure}
Figure \ref{fig: chicken-embryo-data-densities} displays the loglikelihood $\ell(m)$, the likelihood ratio $R(\tau)$ for change-points, the histogram of the
grouped data and the  kernel density $\hat f_K$, the  MLE  $\hat f_{\ML}$, the  MHDE $\hat f_{\MHD}$,   the  AMHDE  $\hat f_{\AMHD}$, and the proposed maximum \bernstein likelihood estimate (MBLE) $\hat f_{\mathrm{B}}$. From this figure we see that the proposed MBLE  $\hat f_{\mathrm{B}}$ and the parametric MLE  $\hat f_{\ML}$ are similar and fit the data reasonably. The kernel density is clearly not a good estimate. The AMHDE  $\hat f_{\AMHD}$ seems to have overestimated $f$ at numbers close to 0.
%\begin{comment}
\section{Concluding Remarks}
The proposed density estimate $f_m(t;\hat{\bm p})$ has obviously considerable advantages over the kernel density: (i) It is more efficient than the kernel density because it is an approximate  maximum likelihood estimate; (ii) It is easier to select an optimal model degree $m$ than to select an optimal bandwidth $h$ for the kernel density; (iii)
The proposed density estimate $f_m(t;\hat{\bm p})$ aims at $f_m(t;\bm p_0)$ which is the best approximate of $f$ for each $m$, while the kernel density $\hat f_K$ aims at
$f*K_h$, the convolution of $f$ and $K_h$.

Another significance of this paper is the introduction of the acceptance-rejection argument in proving the asymptotic results where an approximate model is assumed which is new to the knowledge of the author.
%\end{comment}
%\section*{Acknowledgements} The author want to thank Dr. Nan Lin for sharing his PhD thesis for det.
\appendix
%\section{Appendix}
\section{Proofs}
\subsection{Proof of Theorem \ref{thm: approx of poly w pos coeff}}
\begin{proof}
We define $\Lambda_{r}=\Lambda_{r}(\delta, M_0, M_2,$ $\ldots,M_r)$  as the class of
functions $\phi(t)$ on $[0,1]$ whose first $r$ derivatives $\phi^{(i)}$, $i=1,\ldots,r$, exist and are continuous  with the properties
\begin{equation}\label{eq: boundary conditions}
    \delta\le \phi(t)\le M_0,\quad |\phi^{(i)}(t)|\le M_i,\quad 2\le i\le r,\quad 0\le t\le 1,
\end{equation}
for some $\delta>0$, $M_i>0$, $i=0,2,\ldots,r$. A polynomial of degree $m$ with ``positive coefficients''
is defined by \cite{Lorentz-1963-Math-Annalen} as
$\phi_m(t)=\sum_{i=0}^m c_{i}{m\choose i}t^i(1-t)^{m-i}$, where $c_{i}\ge 0$, $i=0,\ldots,m$.
Theorem 1 of \cite{Lorentz-1963-Math-Annalen} proved that
  for given integers $r\ge 0$, $\delta > 0$, and positive constants $M_i \ge 0$, $i=0,2,3,\ldots,r$, then there exists a
constant $C_r = C_r (\delta, M_0, M_2,$ $\ldots,M_r)$ such that for each function $\phi\in \Lambda_r (\delta, M_0, M_2,\ldots,M_r)$
one can find a sequence $\phi_m$, $m = 1,2,\ldots$,  of polynomials with positive coefficients
of degree $m$ such that
\begin{equation}\label{eq: approx of poly w pos coeff}
    |\phi(t)-\phi_m(t)|\le C_r \Delta_m^{r}(t)\omega(\Delta_m(t), \phi^{(r)}),\quad 0\le t\le 1,
\end{equation}
where $\omega (\delta, \phi) = \sup_{|x-y|\leq \delta} |\phi(x)-\phi(y)|\,$.

 Under the conditions of Theorem \ref{thm: approx of poly w pos coeff},
we see that $M_0=\max_t f(t)$, $M_i=\max_{t}|f^{(i)}(t)|$, $i=2,\ldots,r$, are finite and $\omega(\Delta_m(t), f^{(r)})\le 2M_r$.
So by the above result of  \cite{Lorentz-1963-Math-Annalen} we have a sequence $\phi_m(t)=\sum_{i=0}^m c_{i}{m\choose i}t^i(1-t)^{m-i}$, $m = 0, 1,\ldots$,  of polynomials with positive coefficients
of degree $m$ such that
\begin{equation}\label{eq: poly approx w pos coeff}
    |f(t)-\phi_m(t)|\le 2C_r M_r \Delta_m^{r}(t),\quad 0\le t\le 1.
\end{equation}
%where $C(r,\delta,f)=2C_r M_r$ depending on $r, \delta$, and $M_i$'s only.
\begin{comment}
if $m\le 4$ then $\Delta_m(t)=m^{-1}$, otherwise if $m>4$ then
$$\Delta_m(t)=
\left\{
  \begin{array}{ll}
    \sqrt{{t(1-t)}/{m}}, & \hbox{$|t-0.5|\le 0.5\sqrt{1-{4}/{m}}$;} \\
    m^{-1}, & \hbox{elsewhere.}
  \end{array}
\right.$$
\begin{equation}\label{eq: est of Delta(m)}
    \Delta_m(t)\le
\left\{
  \begin{array}{ll}
    m^{-1}, & \hbox{$m\le 4$;} \\
    0.5m^{-1/2}, & \hbox{elsewhere.}
  \end{array}
\right.
\end{equation}
\end{comment}
It is clear that
\begin{equation}\label{eq: est of Delta(m)}
    \Delta_m(t)\le m^{-1/2}.
\end{equation}
%Thus $\int_0^1\Delta_m^{r}(t)dt\le m^{-r/2}$.
Since $\int_0^1f(t)dt=1$, we have, by (\ref{eq: est of Delta(m)}),
\begin{equation}\label{eq: est of sum of ci}
\Big|1-\sum_{i=0}^m c_i/(m+1)\Big|\le 2C_r M_r m^{-r/2}.
\end{equation}
Let $p_{mi}=c_i/\sum_{i=0}^n c_i$, $i=0,\ldots,m$, then $f_m(t)=\sum_{i=0}^m p_{mi}\beta_{mi}(t)$.
It follows easily from (\ref{eq: poly approx w pos coeff}) and (\ref{eq: est of sum of ci}) that  (\ref{eq: approx of f in Cr}) is true.
\end{proof}
\subsection{Proof of Theorem \ref{thm: convergence of EM algorithm}}
We will prove the assertion (i) only. The assertion (ii) can be proved similarly.
\begin{proof}
The matrix of second derivatives of $\ell_R(\bm p)$ is
$$H(\bm p)=\frac{\partial^2 \ell_R(\bm p)}{\partial \bm p\partial\bm p^\tr}=-\sum_{i=1}^{n}\frac{\bm\beta_m(x_i)\bm\beta_m^\tr(x_i)}{\{\sum_{j=0}^mp_{mj}\beta_{mj}(x_i)\}^2}.$$
For any $\bm u=(u_0,\ldots,u_m)^\tr\in R^{m+1}$, as $n\to\infty$, $$\frac{1}{n}\bm u^\tr\frac{\partial^2 \ell_R(\bm p)}{\partial \bm p\partial\bm p^\tr} \bm u\to \E \left\{\frac{\sum_{j=0}^mu_{j}\beta_{mj}(X)}{\{\sum_{j=0}^mp_{mj}\beta_{mj}(X)}\right\}^2.$$
Clearly, $\beta_{m0}(t),\ldots,\beta_{mm}(t)$  are linearly independent nonvanishing functions on [0,1].
So, with probability one, $H(\bm p)$ is negative definite for all $\bm p$ and sufficiently large $n$. By Theorem 4.2 of \cite{Redner-Walker-1984-siam},
as $s\to\infty$, $\hat {\bm p}^{(s)}$ converges to the maximizer of $\ell_R(\bm p)$ which is unique.
\end{proof}
\subsection{Proof of Lemma\ref{lemma: acceptance-rejection argument}}
\begin{proof}
By (\ref{eq: approx of f in Cr}) and (\ref{eq: est of Delta(m)}) we know that under the condition of the lemma $f_m(t)=f_m(t; \bm p_0)$  converges to $f(t)$ at a rate of at least ${\cal O}(m^{-k})$, i.e.,
\begin{equation}\label{eq: approx of fm to f}
f(t)=f_m(t)+{\cal O}(m^{-k}),
\end{equation}
 and, furthermore, since $f(t)\ge\delta$,
%Let $f$ be a density and $f_m$ a parametric density which belongs to an $m$-dimensional space such that
\begin{equation}\label{eq: max ratio of fm to f}
    c_m=\sup_{t}\frac{f_m(t)}{f(t)}=1+{\cal O}(m^{-k}),
\end{equation}
 uniformly in $m$.

%Assume that $x_1,\ldots,x_n$ is a random sample from $f$.
%We define the approximate loglikelihood $$\ell(f_m)=\sum_{i=1}^n\log f_m(x_i).$$
%We use the argument for generating pseudorandom numbers with acceptance-rejection method in the following proof.
Let $u_1,\ldots,u_n$ be a sample from the uniform(0,1).
By the acceptance-rejection method in simulation \citep{Ross-book}, for each $i$, if $u_i\le  {f_m(x_i)}/{c_mf(x_i)}$, then $x_i$ can be treated as if it were from $f_m$.
Assume that the data $x_1,\ldots,x_n$ have been rearranged so that the first $\nu_m$ observations can be treated as if they were from $f_m$.
By the law of iterated logarithm we have
\begin{eqnarray*}
% \nonumber to remove numbering (before each equation)
  \nu_m &=& \sum_{i=1}^nI\left(u_i\le \frac{f_m(x_i)}{cf(x_i)}\right) \\
    &=& n-{\cal O}(nm^{-k})-{\cal O}\left(\sqrt{nm^{-k}\log\log n}\right),\quad a.s..
\end{eqnarray*}
So we have
$$\ell_R(\bm p)=\sum_{i=1}^n\log f_m(x_i; \bm p)=\tilde\ell_R(\bm p)+R_{mn},$$
where $\tilde\ell_R(\bm p)=\sum_{i=1}^{\nu_m}\log f_m(x_i; \bm p)$ is an ``almost complete'' likelihood
and
$$R_{mn}=\sum_{u_i> \frac{f_m(x_i)}{cf(x_i)}}\log f_m(x_i; \bm p)=\sum_{i=\nu_m+1}^n\log f_m(x_i; \bm p).$$
Because $0<\delta\le f(t)\le M_0$, we have $0<\delta'\le f_m(x_i; \bm p)\le M_0'$ for some constants $\delta'$ and $M_0'$. By the law of iterated logarithm
\begin{eqnarray}\nonumber
% \nonumber to remove numbering (before each equation)
  |R_{mn}| &=& \max\{|\log\delta'|,|\log M_0'|\}\sum_{i=1}^nI\left(u_i> \frac{f_m(x_i)}{c_mf(x_i)}\right)
 \\\nonumber
    &=& \max\{|\log\delta'|,|\log M_0'|\}(n-\nu_m)\\\label{eq: remainder for ell}
&=& {\cal O}(nm^{-k})+{\cal O}\left(\sqrt{nm^{-k}\log\log n}\right),\quad a.s..
\end{eqnarray}
 The proportion of the observations that can be treated as if they were from $f_m$
is $$\frac{\nu_m}{n}=1-{\cal O}(m^{-k})-{\cal O}\left(\sqrt{m^{-k}\log\log n/n}\right),\quad a.s.. $$
\added{So the complete data $x_1,\ldots,x_n$ can be viewed as a slightly  contaminated sample from $f_m$. }
\end{proof}
\subsection{Proof of Theorem \ref{thm: convergence rate for raw data}}
\begin{proof}

The Taylor expansions of $\log {f_m(x_j,\bm p)}$  at $\log {f_m(x_j,\bm p_0)}$ yield  that, for $\bm p\in  \mathbb{B}_m(r_n)$,
\begin{align*}
\tilde\ell_R(\bm{p})&=\sum_{j=1}^{\nu_m} \log {f_m(x_j,\bm p)} \\
&=\tilde\ell_R(\bm{p}_0)+\sum_{j=1}^{\nu_m}\left[\frac{f_m(x_j,\bm p)-f_m(x_j,\bm p_0)}{f_m(x_j,\bm p_0)}-\frac{1}{2}\frac{\{f_m(x_j,\bm p)-f_m(x_j,\bm p_0)\}^2}{\{f_m(x_j,\bm p_0)\}^2}\right]+\tilde R_{mn},%+o(n^{1-2r}).
\end{align*}
where $\tilde R_{mn}=o(nr_n^2)$, a.s..
%$$R_{mn}= %\frac{1}{6}\sum_{j=1}^{\nu_m}\frac{\{f_m(x_j,\bm p)-f_m(x_j,\bm p_0)\}^3}{\{f_m(x_j,\bm p^*)\}^3}={\cal O}(n^{1-3r})
%o(n^{1-2r}),\quad a.s..$$

Let $\bm p$ be a point on the boundary of $\mathbb{B}_m(r_n)$, i.e., $\Vert \bm p-\bm p_0\Vert^2_R=r_n^2$.
By the
law of iterated logarithm we have
\begin{align*}
%\frac{1}{\nu_m}
\sum_{j=1}^{\nu_m}\frac{f_m(x_j,\bm p)-f_m(x_j,\bm p_0)}{f_m(x_j,\bm p_0)}
&=\mathcal{O}(r_n\sqrt{n\log\log n}),\; a.s.,
\end{align*}
and that there exists $\eta>0$ such that
\begin{align*}
%\frac{1}{\nu_m}
\sum_{j=1}^{\nu_m}\frac{\{f_m(x_j,\bm p)-f_m(x_j,\bm p_0)\}^2}{\{f_m(x_j,\bm p_0)\}^2}
%&\ge\frac{1}{M_0\nu_m} \sum_{j=1}^{\nu_m}\frac{\{f_m(x_j,\bm p)-f_m(x_j,\bm p_0)\}^2}{f_m(x_j,\bm p_0)}
%\\
&=\eta nr_n^2+\mathcal{O}(r_n^2\sqrt{n\log\log n}).
\end{align*}
Therefore we have
\begin{align*}
\tilde\ell_R(\bm{p})
&=\tilde\ell_R(\bm{p}_0)+\sum_{j=1}^{\nu_m}\left[\frac{f_m(x_j,\bm p)-f_m(x_j,\bm p_0)}{f_m(x_j,\bm p_0)}-\frac{1}{2}\frac{\{f_m(x_j,\bm p)-f_m(x_j,\bm p_0)\}^2}{\{f_m(x_j,\bm p_0)\}^2}\right]+o(nr_n^2)%{\cal O}(n^{1-3r})
\\
&=\tilde\ell_R(\bm{p}_0)- \frac{1}{2}\eta nr_n^2+\mathcal{O}(r_n^2\sqrt{n\log\log n})+\mathcal{O}(r_n\sqrt{n\log\log n})+o(nr_n^2),\quad a.s..%{\cal O}(n^{1-3r})
%\\
%&\le \tilde\ell(\bm{p}_0)- \frac{1}{2}\eta   n^{1-2r}+\mathcal{O}(n^{-r}\sqrt{n\log\log n})+o(nr_n^2).%{\cal O}(n^{1-3r}).
\end{align*}
%If $r=1/3$,  then
%If $r_n^2=n^{\epsilon-1}$, $0<\epsilon<1$, if $m=Cn^{1/k}$
Since $m={\cal O}(n^{1/k})$,   $nm^{-k} =o(nr_n^2)$.
So there exists $\eta'>0$ such that
$\ell_R(\bm{p})\le \ell_R(\bm{p}_0)- \eta' nr_n^2=\ell_R(\bm{p}_0)- \eta' (\log n)^2$.
Since ${\partial^2 \ell_B(\bm{p})}/{\partial
\bm{p}\partial{\bm p}^\tr}< 0$,  the maximum value of $\ell_R(\bm{p})$ is attained by some $f_m(\cdot,\hat{\bm{p}}_R)$ with $\hat{\bm p}_R$ being in the interior of $\mathbb{B}_m(r_n)$.
%In fact, we can choose $0<r<1/2$.

%If $r_n^2=n^{-1}\log n$,  $m=Cn^{1/k}$ so that  $nm^{-k}=C^{-k}=o(nr_n^2)$.
%We can also choose $r_n^2=(\log n)^2/n$.
\end{proof}
\subsection{Proof of Theorem \ref{cor: mise for raw data}}
\begin{proof}
It is easy to see that (\ref{eq: weighted mise for raw data}) and (\ref{eq: mise for raw data}) follow from  Theorem \ref{thm: convergence rate for raw data}, (\ref{eq: max ratio of fm to f}), the boundedness of $f$, and the triangular inequality.
\end{proof}
\subsection{Proof of Theorem \ref{thm: convergence rate for grouped data}}
\begin{proof}
%Let $f$ be a density and $f_m$ a parametric density which belongs to a $m$-dimensional space such that
By (\ref{eq: approx of fm to f}) we have
\begin{equation}\label{eq: approx of thetam to theta}
    \theta_i=\theta_{mi}(\bm p_0)+{\cal O}(m^{-k}\Delta t_i),
\end{equation}
where
$$\theta_i=\int_{t_{i-1}}^{t_i}f(x)dx,\quad \theta_{mi}(\bm p_0)=\int_{t_{i-1}}^{t_i}f_m(x; \bm p_0)dx,\quad i=1,\ldots,N.$$
Because  $0<\delta\le f(t)\le M_0$  we have
\begin{equation}\label{eq: max ratio of thetam to theta}
    c=\max_{1\le i\le N}\frac{\theta_{mi}(\bm p_0)}{\theta_i}=1+{\cal O}(m^{-k}),
\end{equation}
 uniformly in $m$.  Assume that $y_1,\ldots,y_n$ is a random sample from the discrete distribution with probability mass function $\theta_i=P(Y=i)$, $i=1,\ldots,N$.
%We define the approximate loglikelihood
%$$\ell(\bm\theta_m)=\sum_{i=1}^Nn_i\log \theta_{mi},$$

%We use the argument for generating pseudorandom numbers with acceptance-rejection method in the following proof.
Let $u_1,\ldots,u_n$ be a sample from the uniform(0,1).
For $y_i\sim \bm\theta\equiv(\theta_1,\ldots,\theta_N)$, $i=1,\ldots,n$,  let $u_i\sim U(0,1)$. If $u_i\le c^{-1}{\theta_{my_i}^{(0)}}/{\theta_{yi}}$, then $y_i$ can be treated as if it were from $\bm\theta_m(\bm p_0)\equiv (\theta_{m1}(\bm p_0),\ldots,\theta_{mN}(\bm p_0))$.
Assume that the data $y_1,\ldots,y_n$ have been rearranged so that the first $\nu_m$ observations can be treated as if they were from $\bm\theta_m(\bm p_0)$.
By the law of iterated logarithm we have
\begin{eqnarray*}
% \nonumber to remove numbering (before each equation)
  \nu_m &=& \sum_{i=1}^nI\left(u_i\le \frac{\theta_{my_i}(\bm p_0)}{c\theta_{y_i}}\right)  \\
    &=& n-{\cal O}(nm^{-k})-{\cal O}\left(\sqrt{nm^{-k}\log\log n}\right),\quad a.s..
\end{eqnarray*}
So we have
\begin{eqnarray*}
% \nonumber to remove numbering (before each equation)
  \ell_G(\bm p_0) %&=& \sum_{j=1}^n\sum_{i=1}^NI(y_j=i)\log \theta_{mi}(\bm p_0) \\
    &=& \sum_{i=1}^Nn_i\log \theta_{mi}(\bm p_0)%\\
=\tilde\ell_G(\bm p_0)+R_{mn},
\end{eqnarray*}
where
% $\tilde\ell(\bm p_0)$ is the partial likelihood
\begin{eqnarray*}
% \nonumber to remove numbering (before each equation)
\tilde\ell_G(\bm p_0)&=&\sum_{j=1}^{\nu_m}\sum_{i=1}^N I(y_j=i)\log \theta_{mi}(\bm p_0)%\\
=\sum_{j=1}^{\nu_m}\tilde n_i\log \theta_{mi}(\bm p_0),
\end{eqnarray*}
  $n_i=\#\{j: y_j=i\}$,   $\tilde n_i=\sum_{j=1}^{\nu_m}  I(y_j=i)$, $i=1,\ldots,N$,
and
\begin{eqnarray*}
R_{mn}&=&\sum_{u_j> \frac{\theta_{my_j}}{c\theta_{y_j}}}\sum_{i=1}^NI(y_j=i)\log \theta_{mi}%\\
=\sum_{i=1}^N(n_i-\tilde n_i)\log \theta_{mi}.
\end{eqnarray*}
It is clear that there exist $\delta'>0$ and $M_0'>0$ such that $\delta' \Delta t_i\le \theta_{mi}\le M_0' \Delta t_i$. By the law of iterated logarithm,
\begin{eqnarray*}
|R_{mn}|&\le& \max\{|\log\delta' \Delta t_i|,|\log M_0' \Delta t_i|\} (n-\nu_m)\\
&=&{\cal O}(nm^{-k})+{\cal O}\left(\sqrt{nm^{-k}\log\log n}\right),\quad a.s..
\end{eqnarray*}
%\begin{eqnarray}
%% \nonumber to remove numbering (before each equation)
%  \frac{\partial \ell_G(\bm p)}{\partial p_{mj}} &=& \sum_{i=1}^Nn_i\frac{\{\mathcal{B}_{mj}(t_i)-\mathcal{B}_{mj}(t_{i-1})\}}{\theta_{mi}}, \\
%  \frac{\partial^2 \ell_G(\bm p)}{\partial p_{mj}\partial p_{mk}} &=& %-\sum_{i=1}^Nn_i\frac{\{\mathcal{B}_{mj}(t_i)-\mathcal{B}_{mj}(t_{i-1})\}\{\mathcal{B}_{mk}(t_i)-\mathcal{B}_{mk}(t_{i-1})\}}{\theta_{mi}^2}.
%\end{eqnarray}
The Taylor expansions of $\log {\theta_{mi}(\bm p)}$  at $\log {\theta_{mi}(\bm p_0)}$ yield  that, for $\bm p\in\mathbb{B}_m(r_n)$,
\begin{align*}
\tilde\ell_G(\bm{p})&= \sum_{i=1}^N\tilde n_i \log {\theta_{mi}(\bm p)} \\
&=\tilde\ell_G(\bm{p}_0)+\sum_{i=1}^{N}\tilde n_i\left[\frac{\theta_{mi}(\bm p)-\theta_{mi}(\bm p_0)}{\theta_{mi}(\bm p_0)}-\frac{1}{2}\frac{\{\theta_{mi}(\bm p)-\theta_{mi}(\bm p_0)\}^2}{\{\theta_{mi}(\bm p_0)\}^2}\right]+\tilde R_{mn},%+o(n^{1-2r}).
\end{align*}
where $\tilde R_{mn}=o(nr_n^2)$, a.s..
%$$R_{mn}= %\frac{1}{6}\sum_{j=1}^{\nu_m}\frac{\{f_m(x_j,\bm p)-f_m(x_j,\bm p_0)\}^3}{\{f_m(x_j,\bm p^*)\}^3}={\cal O}(n^{1-3r})
%o(n^{1-2r}),\quad a.s..$$

Let $\bm p$ be a point on the boundary of $\mathbb{B}_m(r_n)$, i.e., $\Vert \bm p-\bm p_0\Vert^2_B=r_n^2$.
It follows from the
law of iterated logarithm that there exists $\eta>0$ such that
\begin{align*}
\frac{1}{\nu_m}\sum_{i=1}^{N}\tilde n_i\frac{\{\theta_{mi}(\bm p)-\theta_{mi}(\bm p_0)\}^2}{\{\theta_{mi}(\bm p_0)\}^2}&=\eta nr_n^2+\mathcal{O}(r_n^2\sqrt{\log\log n/n}).
\end{align*}
Therefore
\begin{align*}
\tilde\ell_G(\bm{p})&=\tilde\ell_G(\bm{p}_0)+\sum_{i=1}^{N}\tilde n_i\left[\frac{\theta_{mi}(\bm p)-\theta_{mi}(\bm p_0)}{\theta_{mi}(\bm p_0)}-\frac{1}{2}\frac{\{\theta_{mi}(\bm p)-\theta_{mi}(\bm p_0)\}^2}{\{\theta_{mi}(\bm p_0)\}^2}\right]+o(nr_n^2)%{\cal O}(n^{1-3r})
\\
&=\tilde\ell_G(\bm{p}_0)- \frac{1}{2}\eta nr_n^2+\mathcal{O}(r_n^2\sqrt{\log\log n/n})+\mathcal{O}(r_n\sqrt{n\log\log n})+o(nr_n^2),\quad a.s..%{\cal O}(n^{1-3r})
%\\
%&\le \tilde\ell(\bm{p}_0)- \frac{1}{2}\eta   n^{1-2r}+\mathcal{O}(n^{-r}\sqrt{n\log\log n})+o(nr_n^2).%{\cal O}(n^{1-3r}).
\end{align*}
%If $r=1/3$,  then
%If $r_n^2=n^{\epsilon-1}$, $0<\epsilon<1$, $m=Cn^{1/k}$ so that  $nm^{-k}=C^{-k}=o(nr_n^2)$.
Since $m={\cal O}(n^{1/k})$, $nm^{-k}= o(nr_n^2)$.
So there exists $\eta'>0$ such that
$\ell_G(\bm{p})\le \ell_G(\bm{p}_0)- \eta' nr_n^2=\ell_G(\bm{p}_0)- \eta' (\log n)^2$.
Since ${\partial^2 \ell_G(\bm{p})}/{\partial
\bm{p}\partial{\bm p}^\tr}< 0$,  the maximum value of $\ell_G(\bm{p})$ is attained by some $\tilde{\bm p}_G$  in the interior of $\mathbb{B}_m(r_n)$.
%In fact, we can choose $0<r<1/2$.

%If $r_n^2=n^{-1}\log n$,  $m=Cn^{1/k}$ so that  $nm^{-k}=C^{-k}=o(nr_n^2)$.
%We can also choose $r_n^2=(\log n)^2/n$.
\end{proof}
\subsection{Proof of Theorem \ref{thm: relationship between the two norms}}
\begin{proof} By (\ref{eq: Riemann sum expression}) and the Taylor expansion we have
%$$\Vert \bm p-\bm p_0\Vert^2_R-\Vert\bm p- \bm p_0\Vert_G^2=\sum_{i=1}^{N}\int_{t_{i-1}}^{t_i}\left[\psi_m(t)-\psi_m(t_{mij}^*)\right]dt$$
%$$=\max_{0\le t\le 1} |\psi_m'(t)| {\cal O}(\max_i \Delta t_i).$$
%$$=\Vert \bm p-\bm p_0\Vert^2_B+{\cal O}(\max_i \Delta t_i) \int_0^1 |\psi_m'(t)|dt .$$
\begin{eqnarray}\nonumber
% \nonumber to remove numbering (before each equation)
  \Vert \bm p-\bm p_0\Vert^2_R %&=& \sum_{i=1}^{N}\int_{t_{i-1}}^{t_i}\left[\psi_m(t)-\psi_m(t_{mij}^*)\right]dt \\
    &=&\Vert\bm p- \bm p_0\Vert_G^2+ {\cal O}(\max_i \Delta t_i) \int_0^1 |\psi_m'(t)|dt\\\label{eq: difference btwn the two norms}
&&\hspace{3em}+\max_{0\le t\le 1} |\psi_m''(t)| {\cal O}(\max_i \Delta t_i^2).
\end{eqnarray}
By  $\beta'_{mj}(t)=(m+1)\{\beta_{m-1,j-1}(t)-\beta_{m-1,j}(t)\}$, we have
$$|\psi_m'(t)|\le m^2 \Big\{ C_1 \sqrt{\psi_m(t)} + C_2 \psi_m(t) \Big\}.$$
It follows easily from  $\beta''_{mj}(t)=m(m+1)\{\beta_{m-2,j-2}(t)-2\beta_{m-2,j-1}(t)+\beta_{m-2,j}(t)\}$ that
$|\psi_m''(t)|\le C_3m^4.$ Thus by (\ref{eq: difference btwn the two norms}) we can obtain
$$\Vert \bm p-\bm p_0\Vert^2_R-\Vert\bm p- \bm p_0\Vert_G^2
%=\sum_{i=1}^{N}\int_{t_{i-1}}^{t_i}\left[\psi_m(t)-\psi_m(t_{mij}^*)\right]dt$$
%$$=\max_{0\le t\le 1} |\psi_m'(t)| {\cal O}(\max_i \Delta t_i).$$
%$$={\cal O}(\max_i \Delta t_i) \int_0^1 |\psi_m'(t)|dt+\max_{0\le t\le 1} |\psi_m''(t)| {\cal O}(\max_i \Delta t_i^2).$$
%$$
={\cal O}(\max_i \Delta t_i){\cal O}(m^2)\Vert \bm p-\bm p_0\Vert_R+{\cal O}(m^4) {\cal O}(\max_i \Delta t_i^2).$$
Therefore we have
$\Vert \bm p-\bm p_0\Vert^2_R=\Vert\bm p- \bm p_0\Vert_G^2+{\cal O}(m^4 \max_i \Delta t_i^2).$
The proof is complete.
\end{proof}
\subsection{Proof of Theorem \ref{cor: mise for grouped data}}
\begin{proof} Similar to the proof of Theorem \ref{cor: mise for raw data},
(\ref{eq: weighted mise for grouped data}) and (\ref{eq: mise for grouped data}) follow  easily from  Theorems \ref{thm: convergence rate for grouped data} and \ref{thm: relationship between the two norms}, (\ref{eq: max ratio of fm to f}), the boundedness of $f$, and the triangular inequality.
\end{proof}
%% HERE WE DECLARE THE BIBLIOGRAPHYSTYLE TO USE AND THE BIBLIOGRAPHY DATABASE
%\bibliographystyle{jasa}
%\bibliographystyle{imsart-nameyear}
%\bibliographystyle{biometrika}
%\bibliography{por-model,bernstein,CHPT,fdr}
\def\polhk#1{\setbox0=\hbox{#1}{\ooalign{\hidewidth
  \lower1.5ex\hbox{`}\hidewidth\crcr\unhbox0}}}
  \def\lfhook#1{\setbox0=\hbox{#1}{\ooalign{\hidewidth
  \lower1.5ex\hbox{'}\hidewidth\crcr\unhbox0}}} \def\cprime{$'$}
  \def\cprime{$'$}

\end{document}